\documentstyle[prl,aps]{revtex}
\draft

\begin{document}
\date{\today }
\title{``Chain scenario'' for Josephson tunneling with $\pi $-shift in
YBa$_2$Cu$_3$%
O$_7$ }
\author{I.I.Mazin$^{(a)}$, A.A. Golubov$^{(b)}$\cite{IFTT}, and A.D.
Zaikin$^{(c)}$%
\cite{FIAN}}
\address{$^{(a)}$Geophysical Laboratory, Carnegie Institution of
Washington, 5251
Broad Branch Road, NW, Washington, DC 20015\\
$^{(b)}$ISI, Research Centre
J\"{u}lich (KFA), FRG\\
$^{(c)}$Institut f\"{u}r Theoretische
Festk\"{o}rperphysik, Universit\"{a}t Karlsruhe, 76128 Karlsruhe, FRG}

\twocolumn[
\maketitle
\begin{abstract}
We point out that all current Josephson-junction experiments
probing directly the symmetry of the
superconducting state in YBa$_2$Cu$_3$O$_7$, can be interpreted
 in terms of the bilayer
antiferromagnetic spin fluctuation model, which renders
the superconducting state with
the order parameters of extended $s$ symmetry, but with the
opposite signs in the
bonding and antibonding Cu-O plane bands. The
essential part of our interpretation
includes the Cu-O chain band which would have
 the order parameter of the same sign
as antibonding plane band. We show that in this case net
Josephson currents along and
perpendicular to the chains have the phase shift equal to $\pi$.
\end{abstract}
\pacs{}] \narrowtext
In the last year, starting with the pioneering work of Wollman et al\cite
{illi}, substantial progress has been made \cite
{jap,eth,dynes,Chaud,ibm2,mathai} in probing the symmetry of superconducting
state in YBa$_2$Cu$_3$O$_7$ by mean of Josephson tunneling. In all these
experiments except of \cite{dynes} the relative phase of the tunneling
currents in YBCO contacts parallel to $a$ and to $b$ crystallographic axes
was measured. In most cases it was found that the phases are opposite, as
expected for instance for $d_{x^2-y^2}.$ To the contrary, in Ref. \cite
{dynes} tunneling current parallel to $c$ was measured, which for pure $%
d_{x^2-y^2}$ is expected to vanish \cite{FN}, and non-zero, although small,
value was found.

Interpretation of the experiments \cite{illi,jap,eth,dynes,Chaud,ibm2,mathai}
is additionally obscured by the fact that the only object studied so far has
been orthorhombic YBa$_2$Cu$_3$O$_7,$ where a $d+s$ state is formally
allowed and one can speak only about the weight of $d-$ or $s-$components.
Some authors \cite{levincarbot} suggested that a strong anisotropy of the
Fermi surface can explain the edge-contact experiments even without large $d-
$component. However, the underlying assumption is that the plane electrons
themselves are subject to strong orthorhombic effects, while both in
calculations \cite{kanazawa} and in the experiment\cite{shen} the main
manifestation of orthorhombicity is the presence of the chains, while the
planes themselves remain fairly tetragonal. This fact cannot be neglected
when judging about pairing symmetry (see, {\it e.g.}, Ref.\cite{kresin}).

 In a previous work \cite{gol},
we noticed that if the order parameters (OP) in chain and plane bands had
opposite sign and if the tunneling current along the chains were dominated
by the chain band, this could explain the Josephson experiments in YBCO
(this suggestion has been recently elaborated on by others \cite{comb}).
While in Ref.\cite{gol} a number of reasons have been proposed for the sign
reversal of the OP, none of Refs.\cite{gol,comb} suggested any microscopical
reason for the tunneling current being dominated by chains.

In this Letter we propose another, quantitative ``chain scenario'' for the
above-mentioned Josephson tunneling experiments. This scenario is based on a
recently proposed bilayer antiferromagnetic spin fluctuation model for
superconductivity in YBCO \cite{LMA} where the symmetry of the pairing state
for the plane electrons is such that the bonding and antibonding plane bands
have OP of the opposite signs while angular symmetry is extended $s$. We
will argue that within this model, if the detailed Fermi surface properties
of YBa$_2$Cu$_3$O$_7,$ as calculated in LDA and measured experimentally, are
taken into account, it follows that the net tunneling currents along $a$ and
along $b$ must have opposite signs.

The Fermi surface of YBa$_2$Cu$_3$O$_7$ is believed to consist of four
sheets: Two plane bands, which are bonding ($b$) and antibonding ($a$)
combinations of the individual planes' states (in the calculations they are
well splitted by the energy which ranges from 0.05 to 0.5 eV, depending on
the wave vector), the chain ($c)$ band, and a small pocket formed mainly by
apical oxygen states (which is not discussed here). The Fermi surface of YBa$%
_2$Cu$_3$O$_7,$ as calculated by Andersen et al, \cite{kanazawa} is shown in
Fig. 1. The $c$ band has been detected by positron annihilation technique
and the corresponding sheet of the Fermi surface is in perfect agreement
with LDA calculations. According to calculations\cite{eilat}, this band is
very light, so that its contribution to the total density of states is small
($\sim 15$\%), while its contribution in the plasma frequency $\omega
_{py}^2\propto N(0)v_{Fy}^2$ is considerable ($\sim 50$\%). These finding
are confirmed by the experiment: Maximal Fermi velocity was calculated\cite
{add} to be $\sim 6\times 10^7$ cm/s and corresponds to the point where the
chain Fermi surface crosses the $\Gamma -$Y line. This value agrees well
with the Raman experiments \cite{friedl}. Calculated plasma frequency
anisotropy $\omega _{py}^2/\omega _{px}^2\simeq 1.75$, as discussed in Ref.%
\cite{MD}, is in agreement with the optical and transport measurements.

Band $a$ is, according to calculations, rather heavy. In particular, it has
extended van Hove singularities which have also been discovered
experimentally \cite{shen,Gofron}. Band $b$ is light again. Both $a$ and $b$
bands are nearly tetragonal. Their relative contribution to the normal-state
transport is defined by the partial plasma frequencies (Table I).
Importantly, bands $a$ and $c$ at $q_z=0$ can cross by symmetry, for
instance they are degenerate with $\epsilon =E_F$ at {\bf q}=($\approx
0.8\pi /a,\approx 0.2\pi /b,0)$. For all $q_z\neq 0$ these bands hybridize.
This is the reason for YBCO being the most three-dimensional of all high-$T_c
$ cuprates. An extremal orbit in $q_z=0$ plane, which appears because of the
$a-c$ hybridization, has been seen in de Haas-van Alphen experiments\cite
{dhva}. These facts provide indirect support for the calculations as regards
$a-b$ splitting and $a-c$ hybridization.

Now we make link to the above mentioned model for the superconducting state,
suggested in Ref.\cite{LMA}. The key feature of the model is that the bands
of the different parity, $a$ and $b$, have OP of the opposite signs. The
sign of the OP in band $c$ was not discussed in Ref.\cite{LMA}. Apparently,
because this band hybridizes with band $a$, but not band $b$, one can assume
that $c$ and $a$ have OP of the same sign, while $b$ has OP of the opposite
sign. How can this fact manifest itself in Josephson tunneling?

To answer this question, let us consider the tunneling currents between two
superconductors, $L$ and $R$, each having several conducting bands, labeled
by subscripts $i,j.$ For simplicity let us assume that the OP $\Delta _i$ in
individulal bands are isotropic. In order to evaluate the Josephson current
between each of these bands we make use of a standard formalism of the
integrated over energy quasiclassical Green functions \cite{Eil} which
should be completed by the boundary conditions \cite{zai} on both sides of
the barrier. Assuming that the tunnel junction transparency is small $D({\bf %
p})\ll 1$ one can proceed perturbatively and expand the boundary conditions
\cite{zai} in powers of $D$. Keeping only the linear terms one gets
\[
g_{Li}({\bf p})-g_{Li}({\bf -p})=\frac 12D_{ij}(f_{Rj}({\bf p})f_{Li}^{+}(%
{\bf p})-f_{Li}({\bf p})f_{Rj}^{+}({\bf p})),
\]
where the functions $g$ and $f$ are respectively the normal and anomalous
quasiclassical Green functions on the left- or on the righthand side of the
barrier. As we are interested only in the linear in $D({\bf p})$
contribution to the Josephson current it is sufficient to substitute for the
anomalous Green functions the unperturbed values, $f_{Li(Rj)}=\Delta
_{Li(Rj)}/\sqrt{|\Delta _{Li(Rj)}|^2+\omega _m^2}$. Then, using $${\bf J}%
=-(2\pi )^{-2}ieT\sum_m\int (d^2S/v_F){\bf v}_F({\bf p})g({\bf p},\omega _m),
$$we arrive at the general expression for the Josephson current between the
bands $i$ and $j$ ($J_{tot}=\sum_{ij}J_{ij}$):
\begin{equation}
J_{ij}=\frac{\pi T}{eR_{ij}}\sum_{\omega _m}\frac{\Delta _{Li}\Delta _{Rj}}{%
\sqrt{|\Delta _{Li}|^2+\omega _m^2}\sqrt{|\Delta _{Rj}|^2+\omega _m^2}}.
\label{4}
\end{equation}

Here $R_{ij}$ is the normal state resistances of a tunnel junction for the
bands $(i,j)$, and is the maximal{\bf \ }of the two resistances, $R_L$ and $%
R_R:$
\begin{equation}
R_{L(R)ij}^{-1}=2e^2\int_{v_x>0}D_{ij}v_{n,Li(Rj)}\frac{d^2S_{Li(Rj)}}{(2\pi
)^3v_{F,Li(Rj)}},  \label{RN}
\end{equation}
$v_n$ is the projection of the Fermi velocity $v_F$ on the direction normal
to the junction plane, $dS$ is an element of the Fermi surface for the
corresponding band. Eq. \ref{4} is a straightforward generalization of the
well known Ambegaokar-Baratoff result \cite{AB} to the case of several
conducting bands.

Further simplification of Eq. \ref{4}. takes place at low temperatures $T\ll
T_c$
\begin{eqnarray}
J_{ij} &=&\frac{2\Delta _i\Delta _j}{eR_{ij}(|\Delta _i|+|\Delta _j|)}%
K\left( \frac{|\Delta _i|-|\Delta _j|}{|\Delta _i|+|\Delta _j|}\right)
\label{6} \\
&\approx &\Delta _i\log 4|\Delta _j/\Delta _i|/eR_{ij},  \eqnum{$\Delta
_i\ll \Delta _j$} \\
&\approx &\pi (\Delta _i\Delta _j)/eR_{ij}(|\Delta _i|+|\Delta _j|),\text{ }
\eqnum{$\Delta _i\approx \Delta _j$}
\end{eqnarray}
where $K(t)$ is the complete elliptic integral. The first case corresponds
to a contact YBa$_2$Cu$_3$O$_7-$ conventional superconductor, and the second
to a grain boundary contact. $\ $

Estimate of the effective barrier transparencies $D_{ij}$ in Eq.\ref{RN} can
be obtained for certain models of the potential barrier $U(x)$ between $L$
and $R.$ Let us consider the case of a specular reflecting barrier $%
U(x)=U_0\delta (x-x_0).$ Transmission probability for a quasiparticle from
the band $i$ in $L$ to tunnel into the band $j$ in $R$ is found by matching
solutions of Schr\"{o}dinger equations on both sides of the barrier using
boundary conditions at the barrier \cite{DN}

\begin{eqnarray}
\Psi _L(x_0) &=&\Psi _R(x_0)  \nonumber \\
U_0\Psi _L(x_0) &=&\frac 1{2m_{Li}}\frac{\partial \Psi _L(x_0)}{\partial x}-%
\frac 1{2m_{Lj}}\frac{\partial \Psi _R(x_0)}{\partial x}  \label{dn}
\end{eqnarray}
The second condition is conservation of the probability current $J(x)=-i{\rm %
Im}\left( \Psi ^{*}\partial \Psi /\partial x\right) /2m(x).$ It is important
to note that, as was shown in Refs.\cite{DN} $m_{i,j}$ are effective band
masses of quasiparticles in $L$ and $R$ , and are neither the bare electron
mass $m_0$, nor the masses renormalized by many-body correlation effects,
i.e., essentially the LDA band masses. As a result, effective barrier
transparency coefficient $D_{ij}$ in Eq.\ref{RN} is determined by the band
velocities:

\begin{equation}
D_{ij}=\frac{v_{n,Li}v_{n,Rj}}{(v_{n,Li}+v_{n,Rj}^2)/4+U_0^2}  \label{8}
\end{equation}
In the low transparency limit, $U_0\gg v$, we have $%
D_{ij}=D_0v_{Li,n}v_{Rj,n}$, where $D_0$ is a constant. Thus, for a
conventional superconductor on the righthand side, we have $\Delta _R\ll
\Delta _{a,b,c},$ and:
\begin{eqnarray}
&&J_a:J_b:J_c  \nonumber \\
&\approx &R_a^{-1}\log \left| \frac{\Delta _a}\Delta \right| :R_b^{-1}\log
\left| \frac{\Delta _c}\Delta \right| :R_c^{-1}\log \left| \frac{\Delta _c}%
\Delta \right|   \nonumber \\
&\approx &v_a^{}:v_b:v_c  \label{J}
\end{eqnarray}

 From Eq.\ref{RN} we observe that band $c$ does not contribute into the
tunneling current in the $x$ direction (i.e., perpendicular to the chains),
which is quite natural. Correspondingly, the total current along $x$ is
\begin{eqnarray}
J_x &=&J_a+J_b=|J_a|-|J_b|  \label{J's} \\
J_y &=&J_a+J_b+J_c=|J_a|-|J_b|+|J_c|  \nonumber
\end{eqnarray}
. Substituting the values for $v$ from Table 1 in Eq. \ref{J} we get
\begin{equation}
J_a:J_b:J_c\approx 1:2:2.2  \label{J1}
\end{equation}

Now we observe that $|J_a|<|J_b|$, while $|J_c+J_a|>|J_b|$, {\it unless }$%
|\Delta _a|,$ $|\Delta _b|$, and $|\Delta _c|$ differ drastically (so that
the log terms in Eq.\ref{J} become important). To check this possibility,
let us come back to the bilayer antiferromagnetic spin fluctuation model of
Ref.\cite{LMA}. The essence of the model is that the coupling interaction is
interband $a-b$ interaction. A non-essential feature of the model was that
the densities of states in both bands were assumed equal. To estimate the
effect of $N_a$ being about twice larger that $N_b$, let us consider the
weak coupling limit for a BCS superconductor near $T_c$ with interband
interaction only. In this limit,
\begin{equation}
\Delta _a=const\cdot V_{ab}N_b\Delta _b\qquad \Delta _b=const\cdot
V_{ab}N_a\Delta _a,  \label{gaps}
\end{equation}
where $V$ is the pairing interaction, and we obtain for the ratio of the
gaps $|\Delta _a/\Delta _b|=\sqrt{N_b/N_a},$ i.e., counterintuitively, the
band with the {\it smaller} density of states, in our case, bonding band,
develops a {\it larger} gap. Including $\Delta _c$ in Eqs.\ref{gaps}
assuming $N_c\ll N_a$, $V_{bc}\approx V_{cc}\approx 0,$ we obtain $|\Delta
_c/\Delta _b|=V_{ac}/V_{ab}.$

Thus, we can safely exclude the possibility of $|\Delta _b|$ being too
small, but there are no arguments within the chosen model that $\Delta _c$
cannot be arbitrary small. However, there are{\bf \ }various experimental
indications that this is not the case, for instance optical experiments of
Bauer et al \cite{genzel}, who compared normal/superconducting optical
conductivity ratios for pure YBCO and YBCO doped with Fe (which substitute
Cu in the chains).{\bf \ }They found that the main effect of doping was that
a gaplike structures at $\omega \sim 150$\ cm$^{-1}$\ shifts down to $\sim 50
$\ cm$^{-1}$, while the maximal gap at $\omega \sim 300$\ cm$^{-1}$\ does
not change. The lower energy can be naturally interpreted as the chain gap
and the higher one as the plane gap. Thus, one can be confident that
relations between $J_a,$ $J_b$, and $J_c$ indeed hold.

Now, since $J_c$ and $J_a$ are in-phase, and $J_b$ is out-of-phase with both
of them, it becomes obvious that the net current along $x$ and along $y$
should have a phase shift $\pi $. Such a state is indistinguishable from $%
d_{x^2-y^2}$ state for those experiments which probe phase difference for
two edge contacts; however, for the tunnel current perpendicular to the
planes the model correctly gives a non-zero value.

The discussion above is relevant for the experiments like Refs. \cite
{jap,eth,mathai}, which deal with YBCO-conventional superconductor contacts.
Let us now discuss the case of YBCO-YBCO contacts \cite{ibm2}
(grain-boundary junctions). The condition for a $\pi $-contact is now $%
|J_{ac}|-|J_{bc}|+|J_{aa}|+|J_{bb}|-2|J_{ab}|<0,$ analogous to Eq.\ref{J's}.
According to Eqs. \ref{6},\ref{8}, the corresponding currents are $%
J_{ij}\propto v_iv_j\Delta _i\Delta _j/(|\Delta _i|+|\Delta _j|).$ To
estimate currents, let us use the relations between the order parameters: $%
\Delta _a/\Delta _b=\sqrt{N_b/N_a}$ and introduce $|\Delta _c/\Delta
_b|=V_{ac}/V_{ab}=\alpha .$ Then the condition is

\[
\frac{\alpha v_av_c\sqrt{N_b}}{\sqrt{N_b}+\alpha \sqrt{N_a}}-\frac{\alpha
v_bv_c}{1+\alpha }+\frac{v_b^2}2+\frac{v_a^2\sqrt{N_b}}{2\sqrt{N_a}}-\frac{%
2v_av_b\sqrt{N_b}}{\sqrt{N_a}+\sqrt{N_b}}<0.
\]
Substituting data from Table I, $v_a:v_b:v_c\sim 1:2:2,$ $N_a:N_b:N_c\sim
2:1:1$ one finds that the above condition holds when $\alpha \agt0.45,$ in
other words, when chain gap is at least half the maximal gap. As discussed
above, a reasonable estimate of the ratio $\alpha $ is close to one half, so
it looks like the condition for existing of the $\pi $-shifts in
grain-boundary junctions is barely satisfied. However, in this kind of
experiments the effect must be very sensitive to the value of the chain gap.
We will return to this issue later.

Our final point concerns the experiments on twinned samples. Naively, one
can assume that the OP in the chain bands have the same sign in all domains,
and thus the tunneling currents from different domains cancel. It easy to
see, however, that it is not the case. Superconducting state in each domain
is degenerate with respect to changing signs of the OP in all bands
simultaneously. Thus the relative phase of the OP in neighboring domains
will be set by the proximity effects. Arguments similar to those used above
for the tunneling currents lead to the conclusion that OP in two adjacent
domains with opposite orientations will have opposite signs of the OP in the
same bands, thus maintaining the proper $a/b$ asymmetry for the whole
crystal.

To summarize, we suggest that recent Josephson junctions experiments that
discovered $\pi $ phase shift between the tunneling currents in $a$ and $b$
directions in YBCO can be fully understood in terms of the $s^{\pm }$
pairing symmetry, when order parameters in bonding and antibonding plane
bands have opposite sign, provided that the chain band is properly taken
into account. This model is able to explain non-zero tunneling current
perpendicular to the planes, as well as independence of the experimental
results on twinning.

A ``smoking gun'' for this model would be an experiment on YBCO with the
superconductivity in the chains intentionally destroyed by doping at the Cu
sites (Fe, Ga), which should be compatible with the conventional $s$%
-pairing. A particularly interesting property of our ``chain scenario'' is
that the experiments with the YBCO-conventional superconductor junctions\cite
{illi,jap,eth,mathai} should be substantially less sensitive to such doping
than the experiments with the grain boundary junctions \cite{ibm2}, since in
the latter case the condition on the chain gap is much more severe.
Interestingly, the only tunneling experiment on YBCO which indeed showed no $%
\pi $-shifts was that of \cite{Chaud}, which was using grain boundary
junctions. This fact is a strong argument in favor of suggested model.

\begin{table}[tbp]
\caption{Partial contributions of the chain, plane- bonding and
plane-antibonding bands to the density of states and plasma frequencies of
YBa$_2$Cu$_3$O$_7$ (from Ref.{\protect\cite{eilat}}).}
\label{N}
\begin{tabular}{l|c|cc|cc}
$n$ & ${\frac{N(n)}{N}}$ & $\left({\frac{\omega_{pl}(n)}{\omega_{pl}}}%
\right)_x$ & $\left({\frac{\omega_{pl}(n)}{\omega_{pl}}}\right)_y$ & $\left({%
\frac{v(n)^2}{v^2}}\right)_x$ & $\left({\frac{v(n)^2}{v^2}}\right)_y$ \\
\tableline bonding & 23\% & 60\% & 37\% & 77\% & 40\% \\
antibonding & 55\% & 37\% & 15\% & 19\% & 7\% \\
chain & 22\% & 3\% & 47\% & 4\% & 54\% \\
&  &  &  &  &
\end{tabular}
\end{table}
\figure{Fermi surface of YBa$_2$Cu$_3$O$_7$, from Ref. {\cite{kanazawa}}}

\end{document}